\documentclass[aps,pra,showpacs,floatfix,superscriptaddress]{revtex4}
\usepackage{graphicx}
\usepackage{subfigure}
\usepackage{color}
\usepackage{ulem}
\usepackage{latexsym}
\usepackage{textcomp}
\usepackage{amsmath}
\usepackage{amssymb}
\usepackage{amsthm}
\usepackage{dsfont}
\begin{document}

\title{ Tomography, Control and  Characterization of Entanglement in  Three level Atomic System}

\author{S. N. Sandhya} 
\email{sns@iitk.ac.in}
\affiliation{Indian Institute of Technology, Kanpur, 208016, India} 
\author{V. Ravishankar} 
\email{vravi@iitk.ac.in}
\affiliation{Indian Institute of Technology, Kanpur, 208016, India}

\begin{abstract}
We study the quantum correlations of the radiation emitted by three level atoms (cascade type) interacting with two driving fields. In the linear regime, and in 
the  Weisskopf-Wigner approximation, we show that the atomic and the two-photon density matrix are equivalent to each other.   This facilitates the 
tomography of the two mode state to be realized  by  measurements on either  the atomic  system or  the emitted fields. While, in general, one needs $4^N$
 measurements for the tomography of a $N$ photon state,  we show that one needs $(N+1)^2-1$ observables  for the tomography of photons emitted by an atomic system. 
 Thus there is an exponential reduction in the  number of observables  for the reconstruction of the class of $N$ photon states emitted by  atoms. 
 We  show that the  driving field strengths and detunings  provide the {\it control} parameters for the preparation  of a specific target state. Finally, we  study the 
 characterization of entanglement of the two photon state. We observe that  a characterization of entanglement in terms of a single  parameter is not possible 
 when the system is in  a mixed state;  therefore, we provide  a description in terms of the newly introduced probability distribution for entanglement, in various regimes of interest.

\end{abstract}


\maketitle

\section{Introduction}
Quantum state tomography (QST) \cite{reviews} is the first step in any  quantum information processing (QIP). 
For a photon qubit, the state determination may be in any of the degrees of freedom,  eg. polarization,
phase space distribution,   or occupancy in\ the Fock space  \cite{reviews}. At the experimental end,  reconstruction of the quantum state  is either by direct 
photon counting measurements or indirectly by optical homodyne /heterodyne  measurements -- as in the case of continuous variable tomography. 
James et al  \cite{Kwait_pra} have proposed a general scheme for  the reconstruction of the polarization state of a pair of entangled photons generated in 
type-II parametric down-conversion (PDC). 
The PDC   two-photon probability distribution  has also been determined experimentally  \cite{kumar} by measuring the joint photon probability distribution.
Theoretically,  Vogel and Risken \cite{vogel} have shown  that the Wigner function can be determined by optical homodyne detection (OHD), which was 
subsequently followed  by the experimental realization of the same \cite{raymer}.  
 The  phase-space quasi probability function has also been determined \cite{wodkiewicz} by photon counting measurements. Single photon
\cite{lvovsky1} and two photon Fock state tomography have been experimentally studied by Lvovsky et al \cite{lvovsky2}. Spin state tomography has been 
studied using group theoretic methods \cite{D'ariano}.
So far, tomography studies on photons have been mainly in PDC systems. 

Preparation of a specific target state as required by the protocol of interest is also  necessary for any QIP. Control of 
atomic sources for the realization of specific photon states may be achieved via coherent atom photon interactions
where the driving field strengths and the detunings of the applied fields  are the {\it control} parameters .
An inevitable consequence of atom photon interactions (coherent) is the existence of strong quantum correlations \cite{sns, zubairy, scully}.  Coherent driving
 fields interacting with the atoms give rise to strong atomic coherence which then gets transferred to the emitted radiation.  
Quantum correlations,  particularly  the  entanglement aspects,  have been the most studied but not easily understood, especially when the state is not pure. 
On the other hand,  the preparation of pure entangled states
itself poses a challenge. While PDC sources offer good fidelity the conversion efficiency is, however, very low since it is a second order process. Further, the 
input field strength is the only control parameter.
A further disadvantage is that there is no control over the emission statistics. On the other hand it is well known that the statistics of emission in resonance 
fluorescence can be made sub-poissonian or super-poissonian. In other words, there can be programmable delays between successive emissions. The 
disadvantage of monitoring emission in resonance fluorescence is that the emitted radiation is over $4\pi$ and the signal to noise ratio per solid angle may 
be  small. This may be overcome by coupling to a cavity which is tuned to the mode frequency of emission.

 In this paper we consider  three level cascade systems interacting with two driving fields. We study the quantum correlations of the radiation emitted by these systems in free space. In particular,  we address three issues :
 
 \noindent  {\it (i)  quantum state tomography of the two-mode emitted radiation}: In the far-field approximation (lowest order in $1/r$, and $t> r/c$ ) and using the Weisskopf-Wigner theory  of spontaneous emission, we show that  it is sufficient to measure a set of 8 observables to completely determine the two-photon density matrix .
This is because a $N$ - photon state of the emitted radiation in our system is restricted  to a Hilbert space of dimension $(N+1)^2-1$ instead of the full Hilbert space of  dimension $2^{N}$.  Accordingly, there is an exponential decrease in the number of observables that are needed for tomography. For the present case, $N=2$, and the number of required observables is 8.   An additional feature is  that there is a one-one mapping of the atomic and the photonic density matrix and therefore a knowledge of one leads to the other.  Unlike  conventional methods of tomography where measurements are restricted to the emitted radiation only, in our scheme there is no need to perform the measurements entirely on either the emitted radiation or on the atomic system.  We identify six observables of the atomic system and two observables of the emitted radiation as a convenient set. Thus this approach has an edge over the conventional methods of tomography. 

\noindent {\it (ii) Control and preparation of specific states}: As already mentioned, it is not easy to prepare entangled systems with a high degree of purity, and with high efficiency. We show that three level cascade systems provide a very good source of entangled two mode emissions with very large probabilities. In fact, even in the steady state one has the option of preparing almost pure fiducial  states, Bell states with reasonable fidelity,  or mixed states by suitably setting the driving field strengths and detunings.  
 
 \noindent {\it (iii) Characterization of  mixed state entanglement:}    Conventional entanglement measures such as concurrence and negativity are known to provide an incomplete description of entanglement in the  mixed state regime. We evaluate the probability density of entanglement for the two-mode state following Bhardwaj and Ravishankar \cite{shantanu}. 
 
 The plan of the paper is as follows: In section II we describe the model,  and set up the equation of motion. In section III we demonstarte the reconstructibility of  the atomic density matrix and also the photon density matrix by a suitable set of measurements.
In section IV we discuss the control of pure/mixed state preparation and the characterization of entanglement.

\section{Model}

Emission from three level cascade sytems are of interest,  especially for the strong quantum correlations that they exhibit. The earliest classic work of Clauser \cite{clauser} in Cs atoms demonstrated the non-classicality of the emitted radiation in these sytems.  The resonance fluorescence \cite{kimble} and the emission statistics \cite{loudon} in this system have been well studied and  the behavior is qualitatively predictable in most regions of the parameter space. In view of this,  and the large number of recent experiments on  electromagnetically induced transparency (EIT)  in $^{87}Rb$ vapours \cite{banacloche},  we choose to study  three level cascade systems.  In passing, we would like to mention that the analysis presented in this paper is equally applicable to  $'\Lambda' $and $'V'$
systems by properly taking care of the detunings. 
The scheme, shown in Fig.1, has only two allowed dipole transitions of energy separations $\omega_1$ and $\omega_2$ corresponding to  the $|2\rangle \rightarrow |1 \rangle $ and $|3\rangle  \rightarrow |2 \rangle$  transitions.  Two counter propagating ( Doppler free geometry) driving fields of nearly equal frequencies $\omega_{L1}$ and $\omega_{L2}$ and respective strengths $\Omega_1$ and $\Omega_2$ are resonant with these two transitions. The decay constants of the energy levels  $| 3 \rangle $ and $| 2\rangle $ are indicated by $ \Gamma_3$ and $\Gamma_2$ respectively. The parameters $\Delta_1,~ \Delta_2 $  refer to the detunings of the driving fields.
\begin{figure}
\includegraphics{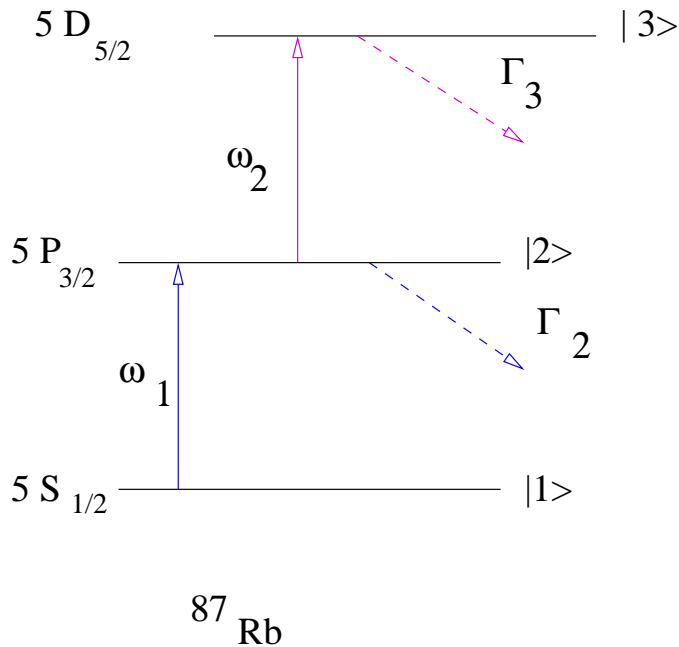}
\caption{(Color online) Three level cascade system corresponding to the $^{87}Rb$ atoms driven by two fields
$\omega_1$ and $\omega_2$.}
\label{fig:1}
\end{figure}

\subsection{Equations of motion}

Since the input parameters are known -- and so is the dynamics --  it is  straightforward to solve for the state of the atomic system.
There is, however,  no corresponding procedure (availability of a master equation) that determines the state of the emitted radiation. One way is  to determine the atom+radiation density matrix, and then trace over the atomic part which is  cumbersome. In this paper we will show that this tedious procedure can be circumvented. We further show that  the atomic density matrix determines the photon density matrix completely!

We start by making an ansatz for the atom+field state as

\begin{equation}
|\Psi (t) \rangle = \sum_{i,\gamma} \alpha_{i}^{\gamma} (t) |i ; \gamma \rangle; i=1, 3; \gamma=0,1, 2, 3;
\end{equation}
where $i$ denotes the atomic index while $\gamma$ denotes the photon mode index. We alternately use the notation  $| i; \gamma \rangle \equiv | i; \gamma_2, \gamma_1 \rangle $. The notation $|i; \gamma \rangle; \gamma = 0,1, 2, 3$ is more commonly used in quantum information where $\gamma = \gamma_2\gamma_1$ in base 2. More explicitly,  $\gamma = 2^0\gamma_1  + 2^1\gamma_2$. In writing the above equation we make the following assumptions which are central to our analysis: (i) 
The rotating wave approximation (RWA) which restricts the transitions  such that a photon is created only when an atomic excitation is annihilated and vice- versa. This is very similar to the ansatz of   Weisskopf -Wigner theory of spontaneous emission. (ii) The  single-photon approximation \cite{single} - the rise time or the excitation time of the atom is inversely proportional to the driving field strength. When $\Omega$ is small compared to the atomic decay constant $\Gamma$, the rise time $1/\Omega$ is large and the atom completes fewer cycles of excitation/de-excitation in an atomic lifetime. This situation enables us to assume that for small times ($t< $ atomic lifetime) the Fock space of the two modes of radiation has at most one photon in each mode. Thus $\gamma_1, \gamma_2=0,1$  only. We therefore restrict our study to the situation where the ground state excitation is weak. 

The interaction Hamiltonian in the interaction picture and in the RWA and the dipole approximation  has the form
\begin{equation}
 H= \frac{\hbar}{2}  ( \omega_1 \sigma^z_{1} + \omega_2 \sigma^z_{2} )+ 
\hbar \Omega_1 (e^{-i \omega_1 t} \sigma^{+}_{1} + h.c.) +  \hbar \Omega_2 (e^{-i \omega_2 t} \sigma^{+}_{2} + h.c.)
\end{equation}
Here we treat the driving fields classically and   $\Omega_i= -\frac{1}{\hbar}\vec {\mu}_{i i+1} \cdot \vec{E}_i;~ i=1,2$, are the Rabi frequencies corresponding to the two driving field strengths.   $\sigma^{+}_{i}= |i+1 \rangle ~\langle i|; ~\sigma^{-}_{i}= |i \rangle ~\langle i+1|; ~ i = 1, 2 $ and $\sigma^z_{i} = |i+1 \rangle ~\langle i+1|- |i \rangle ~\langle i|$ are the  atomic transition operators.
The equation of motion for the 12 complex amplitudes $\alpha_{i}^{\gamma}$ is then given by
\begin{equation}
i \hbar \dot \alpha_{i}^{\gamma} (t)= \langle i; \gamma |   H | \Psi(t) \rangle
\end{equation} 
 It might appear  that, in general,  the two-photon density matrix elements corresponding to the two modes 
\begin{equation}
\rho^{F}_{\gamma \gamma'}=  \sum_i \alpha^{\gamma}_i  \alpha^{\gamma'^*}_{i} 
\end{equation}
would require a set of 16 observables for its full determination.
However,  by employing an equivalent approach in the next section, we show that not all of the complex amplitudes $\alpha_i^{\gamma}(t) $ are independent since they are constrained by the production process and selection rules.  We further  show that $\alpha^1_2,\alpha^1_3,  \alpha^2_1, \alpha^2_3, \alpha^3_2, \alpha^3_3$ are zero  because of RWA . 

\section{Reconstruction of the two-photon state}

\subsection{Equivalence of the atomic and photon density matrix}

 In general, the field 
$\vec E({\bf r},t)$ emitted by the atom has the mode expansion 
\begin{equation}
\vec E^{(+)}({\bf r}, t)= i\sum_{k,\lambda} {\cal E}_k \hat {\epsilon}_{k,\lambda} a_{k, \lambda}(t)e^{i{\bf k}\cdot {\bf r}}
\end{equation}
for the positive frequency part where ${\cal E}_k =(2 \pi \hbar  c k/V)^{1/2}$ for a given mode $k$ and
 $\lambda$ is the polarization index. We now assume that the fields emitted by the two transitions to be dominantly centered around $k_1$ and $k_2$ respectively and are  
defined as in eq(2.12) of Titulaer and Glauber \cite{glauber}.
The creation and annihilation operators satisfy the usual commutation relations
$[\hat a_{k,\lambda} (t), \hat a_{k', \lambda'}^{\dagger} (t)]=  \delta_{k,k'} \delta_{\lambda, \lambda'}$  for a given mode $k$.
In order to establish the equivalence of the atomic state with that of the radiation we consider the auxiliary state in the full Hilbert space of atom+radiation defined by   
\begin{eqnarray}
 {\rho}(t) &=& \sum_i  \sum_{i'}\sum_{p, q, p',q'}B^{i,i^{\prime}}_{p, q; p',q'} (a_{k_2}^{\dagger}(t))^p ( a_{k_1}^{\dagger} (t))^q|i; {0}_{k_2}; {0}_{k_1} \rangle ~\langle i'; {0}_{k_2}; {0}_{k_1}|(a_{k_1}(t))^{q'}( a_{k_2}(t))^{p' }\nonumber \\
\end{eqnarray}
 In the single photon approximation, each mode has either no photons or at most one photon, i.e, $\{p, q, p', q'\}=\{0, 1\}$. The indices  $ k_1, k_2$ correspond to  the two modes and $ |0_{k_2}; 0_{k_1}\rangle $  is the corresponding  vacuum. Recall that  $\rho^{A+F}(t=0)= \rho^A \otimes \rho^F$. Also recall that the  annihilation operators for the two different modes are mutually commuting at a given time, i.e., $ [ a_{k_2}(t) , a_{k_1}(t) ]= \delta_{k_1k_2}$ and likewise for the creation operators. The two-photon density matrix may now be obtained by taking a partial trace over the atomic variables to be
\begin{eqnarray}
\rho^{\gamma}(t ) \equiv \sum_i<i\vert \rho \vert i > =  \sum_{p, q, p',q'}B_{p, q; p',q'} (a_{k_1}^{\dagger}(t))^p ( a_{k_2}^{\dagger} (t))^q| {0}_{k_1}; {0}_{k_2} \rangle ~\langle {0}_{k_1}; {0}_{k_2}|(a_{k_2}(t))^{q'}( a_{k_1}(t))^{p' } \nonumber \\
\end{eqnarray}
where the coefficients $B_{p,q;p',q'} = \sum_iB^{ii}_{p,q;p',q'}$. This expression  is equivalent to the form
that would follow from eq.1, as may be seen by observing that $|i;\gamma\rangle \equiv |i; \gamma_1, \gamma_2 \rangle = (a_2^{\dagger})^{\gamma_2}(a_1^{\dagger})^{\gamma_1}|i; 0_2; 0_1\rangle$.
 Note that this is very similar to the definition of the  most general density operator of n photons in a single mode given by  eq.2.15 in \cite{glauber}. 
Following the Weisskopf-Wigner theory of spontaneous emission, we  stipulate the far field approximation wherein the  field operators $\vec E^{+}({\bf r},t)$ of the emitted radiation at the point of the detection $\bf r$  are  proportional to the  atomic spin operators at the retarded time $\hat \sigma_{i} (t-|{\bf r}-{\bf r}_0|/c) $. In the far field  for an atom located at ${\bf r}_0$,  terms proportional to $O(1/|{\bf r}-{\bf r}_0|^2)$ may be neglected \cite{scully1, mandel}. Thus, for a given mode $k$ of the source field, 
\begin{equation}
\vec E^{(+)}_k( {\bf r},t)= \vec E_0^+({\bf r},t)+ \frac{\eta}{4 \pi \epsilon_0} \frac{\omega_m^2}{c^2} \frac {\vec \mu_{k k+1}} {|{\bf r}_k-{\bf r}_0|} {\hat \sigma}_{k}^-(t-| {\bf r}-{\bf r}_0|/c)
\end{equation}
where $\vec E_0^+({\bf r},t)$ is the vacuum field contribution and $\eta$ is the detector efficiency. 
In this paper we consider the ideal situation of $\eta=1$. The first term due to the vacuum does not contribute to any signal detection and may be ignored \cite{cohen} for detection times larger than the inverse of the optical frequencies. We use the notation $\hat \sigma^-_k= \hat \sigma_{k k+1}$ and  $\hat \sigma^+_k= \hat \sigma_{k+1 k}$. For convenience we replace the mode labels $k_1$ and $k_2$ with '$1$' and '$2$' respectively.  It is convenient to introduce the auxiliary state
\begin{eqnarray}
\bar \rho^{\gamma}(t)&=&\rho^{\gamma} \otimes {\cal I}^A \nonumber \\
&\equiv& f_2( r_2) f_1( r_1) \sum_{i} \sum_{ p, q, p',q'}B_{p, q, p',q'} (\sigma_2^{+}(t_2/c))^p ( \sigma_1^{+} (t_1/c))^q|i; {0}_{2}; {0}_{1} \rangle ~\langle i; {0}_{2}; {0}_{1}|(\sigma_{1}^{-}(t_1/c))^{q'}( \sigma_{2}^{-}(t_2/c))^{p' }
\end{eqnarray}
where $f_k( r_k)= (-\frac{i}{4 \pi \epsilon_0 {\cal E}_k }\frac{\omega_k^2}{c^2} \frac{|\mu_{k k+1}|}{|{\bf r}_k|})^2;~ k=1, 2$,
and $t_i=t-r_i/c ; i=1,2.$  The atomic  operators satisfy the commutation relations 
$ [\sigma_{ij}, \sigma_{kl}]= \delta_{jk} \sigma_{il}- \delta_{li} \sigma_{kj}$  which follow from  the identity                              $ \hat \sigma_{ij}(t) \hat \sigma_{kl}(t)  = \hat \sigma_{il} (t) \delta_{jk}$ for the atomic operators $\sigma_{ij}(0) \equiv |i  \rangle ~\langle j |$.  When the point of observation is the same for both the modes
 ${\bf r}_2={\bf  r}_1= {\bf r}$. Therefore
\begin{equation}
\bar \rho^{\gamma}(t)= f_2(r)f_1(r)  \sum_i \sum_{p, q, p',q'}B_{p, q, p',q'} (\sigma_2^{+}(t- r
/c))^p ( \sigma_1^{+} (t- r/c))^q|i; {0}_{2}; {0}_{1} \rangle ~\langle i; {0}_{2}; {0}_{1}|(\sigma_{1}^{-}(t- r/c))^{q'}( \sigma_{2}^{-}(t- r/c))^{p' }
\end{equation}
Again since the atom and photon density matrices are separable at t=0, the summation term in the r.h.s is nothing but the atomic density matrix. Thus, it follows that 
\begin{equation}
\rho^{\gamma}(t) \otimes {\cal I}^A \rightleftharpoons  {\cal N} \rho^{\gamma}_0 \otimes \rho^A(t-r/c), 
\end{equation}
where $ {\cal N} = f_2(r)f_1(r) $ is the geometric factor and $ ~ \rho^{\gamma}_0= | 0_1; 0_2  \rangle ~\langle 0_1; 0_2 | $. The normalized density matrices are obtained by dropping ${\cal N}$. 

 Eq. 11 merits some more explanation. First of all, $\rho^{\gamma} (t) $ at any time gets determined completely by $\rho^{A} (t- \frac{r}{c})$,  the atomic state at an earlier time $t-r/c$. More interestingly, we observe that  $\rho^{\gamma}$ is of rank three. This reduction in the rank is a manifestation of Schmidt decomposition as applied to the atom+ radiation system. 
More specifically,  the dynamics -- which leads to the equivalence of the atomic (retarded time)  and the field operators, yields the following important relations: (i)  $\rho^{\gamma}$ belongs to the space orthogonal to the state $\vert 2 \rangle_{\gamma}  \equiv \vert 10 \rangle$. (ii) Let us set up  matrix elements for $\rho^{\gamma}$ in the basis  $\{\vert 00 \rangle,~ \vert 01 \rangle,~ \vert 11 \rangle \}$. Eq. 11 asserts that the matrix in this basis  is identically equal to the matrix for the atomic state set up in its standard basis labeled by the energy eigenstates $\{ \vert 1 \rangle,~ \vert 2 \rangle,~ \vert 3 \rangle \}$.  Explicitly,
the normalized photon density matrix has the form

\begin{eqnarray}
\rho^{\gamma}(t)&=&\frac{1}{\cal N}\left (
\begin{array}{ccc}
\sum_i |\alpha^0_i|^2 & \sum_i \alpha^0_i\alpha^{1^*}_i   & \sum_i \alpha^0_i\alpha^{3^*}_i \\
\sum_i \alpha^1_i\alpha^{0^*}_i &\sum_i |\alpha^1_i|^2   & \sum_i \alpha^1_i\alpha^{3^*}_i \\
 \sum_i \alpha^3_i\alpha^{0^*}_i & \sum_i \alpha^3_i\alpha^{1^*}_i  &\sum_i |\alpha^3_i|^2
\end{array}
\right )  \nonumber \\
&\equiv& \left
 ( \begin{array}{ccc}
\rho_{11}^A & \rho_{12}^A &  \rho_{13}^A \\
\rho_{21}^A & \rho_{22}^A &  \rho_{23}^A \\
\rho_{31}^A & \rho_{32}^A &  \rho_{33}^A
\end{array} \right )_{t_r} 
= \rho^A(t-r/c)
\end{eqnarray}
where the subscript $t_r$  emphasizes  that the atomic density matrix is evaluated at a retarded time  $t_r=t-r/c$.  The atomic density matrix may be determined by solving the Liouville equation

\begin{equation}
i \hbar \dot \rho= [ H, \rho] -\frac{i\hbar}{2}\{\Gamma,\rho\}
\end{equation}
where $H$ is the Hamiltonian given by eq(2) and $\Gamma$ is the relaxation matrix. The second term is explicitly given by $- i \hbar \frac{\Gamma_2}{2} \sigma^{+}_{1}\sigma^{-}_{1} - i \hbar \frac{\Gamma_3}{2} \sigma^{+}_{2}\sigma^{-}_{2} $.

\subsection{Measurements and Observables}

One of the important consequences of the equivalence of the atomic and the photon density matrix, shown in Eq. 12,  is that it gives us the freedom of choosing measurements on either the atomic or the photons emitted depending on the practical feasibility. The existence of more than one  complete set of measurements further helps in minimizing uncertainties in the determination of the quantum state. The complex amplitudes,  and hence the density matrix elements, may be determined  by making the simple observation   \cite{lax} that
\begin{equation}
< O^{F}(t) > = Tr_F( O^{F}(t) \rho^{F}) = Tr_{F} (O^{F} (t)Tr_{A} (\rho^{A+F})) \\
=Tr_{A+F}( O^F \otimes {\cal I}^A \rho^{A+F})
\end{equation}
where $F$ and $A$ correspond to the field and atomic variables,  $O^F$ are the  field operators and ${\cal I}^A$ corresponds to the identity operator in the atomic space.  
We consider all expectation values to be normal ordered in the operators. This ensures that the spurious contribution of the vacuum is eliminated. 
The action of the creation and annihilation field  operators  on the atom+field states is given below :
\begin{eqnarray}
 \hat a_{k_2} | i ; k_2; k_1  > & \neq 0& {\rm for} \  k_2=1 \ {\rm and} \  \forall  
\  \{i, k_1 \}; \nonumber \\
 &=0&  \ {\rm for}\  k_2= 0\  {\rm and}\  \forall \  \{i, k_1 \}; 
\end{eqnarray}

The expectation value of the  annihilation operators   
 $\hat a_k; k=1, 2 $ of the two modes is given by
 \begin{eqnarray}
 <\hat  a_{k}(t) > &=& Tr_{A+ F} ( \hat a_{k}(t) \rho^{A+ F}) \nonumber \\
                    &=& \sum_{i, i'} \sum_{k_1, k_1'} \sum_{k_2, k_2'} |i' ; k_2'; k_1' > < i' ; k_2'; k_1' | \nonumber \\
                    & \times&  \hat {\cal O}^{AA'}_{k_2 k_1 k_2'k_1'}(t) |i; k_2; k_1 > <i ; k_2; k_1 | \Psi > < \Psi |
 \end{eqnarray}
where $ \hat {\cal O}^{AA'}_{k_2k_1k_2'k_1'}= \hat {\cal I}_{k_2'} \otimes \hat {\cal I}_{k_1'} \otimes \hat {\cal I}^{A'} \otimes$
 $\hat a_{k_2} \otimes \hat {\cal I}_{k_1} \otimes \hat {\cal I}^A$ and $ \hat {\cal I}_{k_2(k_1)}$ is the identity operator  of mode $k_2(k_1)$ of the field. Identifying $k_2=2, k_1=1$ the expectation values simplify to 
\begin{eqnarray}
<\hat a_1(t)>& = &\sum_i \alpha_i^{1}(t) \alpha_i^{0^*}(t) + \sum_i \alpha^3_i(t) \alpha_i^{2^*}(t) =\alpha_1^{1}(t) \alpha_1^{0^*}(t) \nonumber \\
<\hat a_2(t)>& = &\sum_i \alpha_i^{2}(t) \alpha_i^{0^*} (t)+ \sum_i \alpha^3_i(t) \alpha_i^{1^*}(t)=\alpha_2^{2}(t) \alpha_2^{0^*} (t)+  \alpha^3_1(t) \alpha_1^{1^*}(t)
\end{eqnarray}
since the complex amplitudes $\{\alpha^1_2, \alpha^1_3, \alpha^2_1, \alpha^2_3, \alpha^3_2, \alpha^3_3\} = 0$ due to RWA.  The expectation values of the higher order correlation functions, specifically the normal ordered field correlation and the intensity correlation functions which are nonzero,  are given by
\begin{eqnarray}
< \hat a_1^{\dagger} (t) \hat a_1(t) > &=& \sum_i | \alpha_i^1(t)|^2 = |\alpha_1^1(t)|^2 \nonumber \\
< \hat a_2^{\dagger} (t) \hat a_2(t) >&=& |\alpha_2^2(t)|^2+| \alpha_1^3(t)| ^2 \nonumber \\
<\hat a_2^{\dagger}(t) \hat a_1^{\dagger}(t)>& =&\sum_i \alpha_i^0(t) \alpha_i^{3^*}(t)=  \alpha_1^0(t) \alpha_1^{3^*}(t)\nonumber \\
<\hat a_2^{\dagger}(t) \hat a_1^{\dagger}(t) \hat a_1(t) \hat a_2(t) > &=& | \alpha_1^3 (t) |^2.
\end{eqnarray}

We observe that  $<\hat a_2^{\dagger}(t) \hat a_1^{\dagger}(t)>$ is the anomalous  coherence \cite{gsa} or the cross correlation of the amplitudes of the two fields emitted by the atomic system. On the other hand, the expectation value of the Pauli operators  for the two transitions  is listed below:
\begin{eqnarray}
<\hat \sigma_{-}^{(1)}(t) >& = &\sum_{\gamma} \alpha_2^{\gamma}(t) \alpha_1^{\gamma^*}(t) = \alpha_2^{0}(t) \alpha_1^{0^*}(t) \nonumber \\
<\hat \sigma_{-}^{(2)}(t)>& = &\sum_{\gamma} \alpha^{\gamma}_{3} (t)\alpha^{\gamma^*}_{2}(t) =\alpha^{0}_{3}(t) \alpha^{0^*}_2(t) .
\end{eqnarray}
\begin{figure}
\subfigure[]{\includegraphics[width=8.0cm]{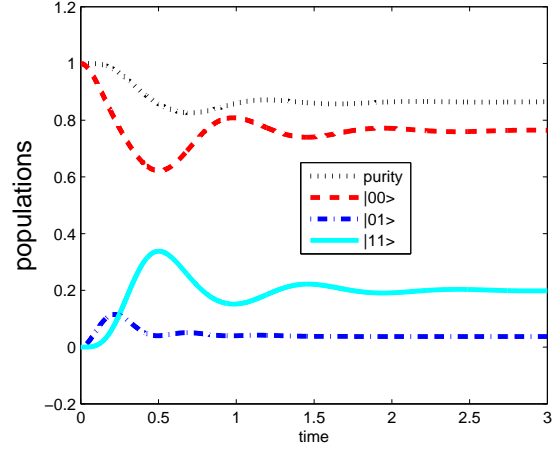}} \\
\subfigure[]{\includegraphics[width=8.0cm]{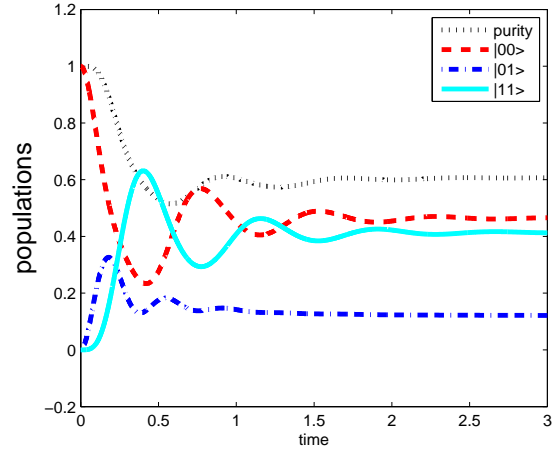}} \\
\subfigure[]{\includegraphics[width=8.0cm]{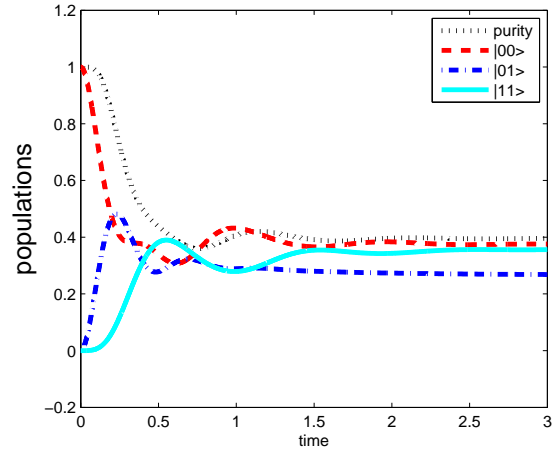}}
\caption{(Color online) Time evolution of the populations for (a) pure state regime, $\Omega_1=3.0, \Omega_2=6.0$
(b) the Bell state regime $\Omega_1=6.0, \Omega_2=6.0~ $ and (c) the mixed state regime $\Omega_1=6.0,~ \Omega_2=3.0$. The black dotted curve corresponds to $Tr \rho^2$ (purity).
}\label{fig:population}
\end{figure}

The expectation values of the normal ordered second and the fourth order spin correlation operators are
given by
\begin{eqnarray}
<\hat \sigma_{+}^{(1)}(t) \hat \sigma_{-}^{(1)}(t) >& = &\sum_{\gamma}|\alpha^{\gamma}_2(t)|^2= |\alpha^{0}_2(t)|^2\nonumber \\
<\hat \sigma_{+}^{(2)}(t) \hat \sigma_{-}^{(2)}(t) > & =&\sum_{\gamma} |\alpha^{\gamma}_3(t)|^2= |\alpha^{0}_3(t)|^2 \nonumber \\
<\hat \sigma_{+}^{(2)}(t) \hat \sigma_{+}^{(1)}(t) > & =&\sum_{\gamma} \alpha^{\gamma}_1(t) \alpha^{\gamma^*}_3(t)= \alpha^{0}_1(t) \alpha^{0^*}_3(t)\nonumber \\
< \hat \sigma_{+}^{(2)}(t) \hat \sigma_{+}^{(1)}(t) \hat \sigma_{-}^{(1)}(t) \hat \sigma_{-}^{(2)}(t) >& =&\sum_{\gamma} |\alpha_3^{\gamma}(t)|^2 = |\alpha_3^{0}(t)|^2.
\end{eqnarray}
From the above equation we see that 
$ <\hat \sigma_{+}^{(2)}(t) \hat \sigma_{+}^{(1)}(t) \hat \sigma_{-}^{(1)}(t) \hat \sigma_{-}^{(2)}(t) >  $
$= <\hat \sigma_{+}^{(2)}(t) \hat \sigma_{-}^{(2)}(t) >$ and the equivalence between the field operators and the atomic transition operators  imply the following set of identities:
$$
|\alpha_2^2(t)|^2= 0; ~ 
 \alpha_3^{0}(t-r/c)  \propto \alpha_1^3 (t);~  \alpha^{0}_2(t-r/c)
 \propto  \alpha_1^1(t).
 $$ 
  The first of the conditions is a manifestation of the Schmidt decomposition which forces $\rho^{\gamma}$ to be of rank three. Note that the atomic operators  
  are related to the  amplitudes
  $\{ \alpha^0_1, \alpha^0_2, \alpha^0_3\} $ only while the emitted radiation is related to  the amplitudes 
   $\{\alpha^0_1, \alpha^1_1, \alpha^3_1\}$ only.
 As already mentioned depending on the practical feasibility, one may choose observables from either the field space or the atomic space. 
 
\begin{figure}
\subfigure[]{\includegraphics[width=7.0cm]{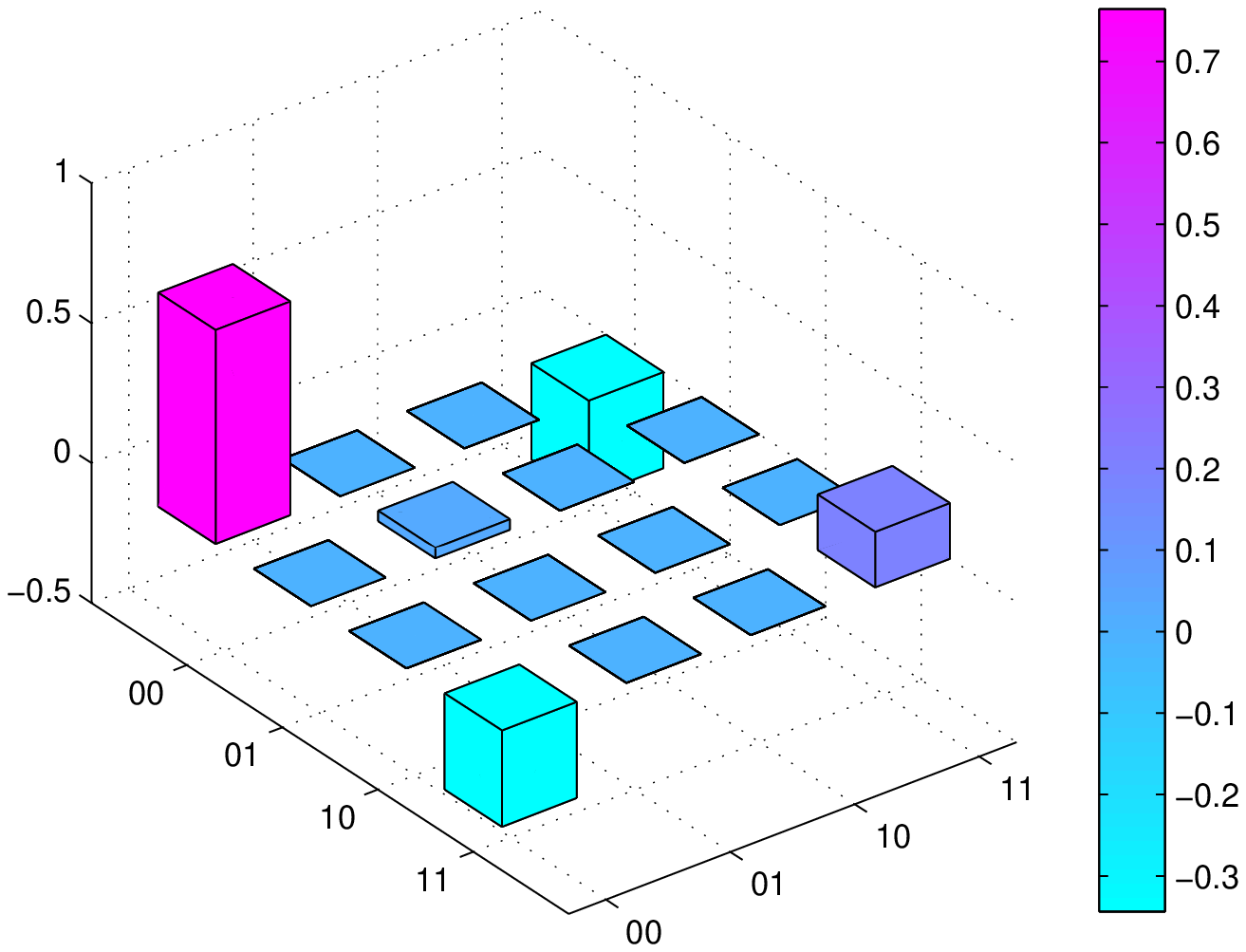}} \hfill
\subfigure[]{\includegraphics[width=7.0cm]{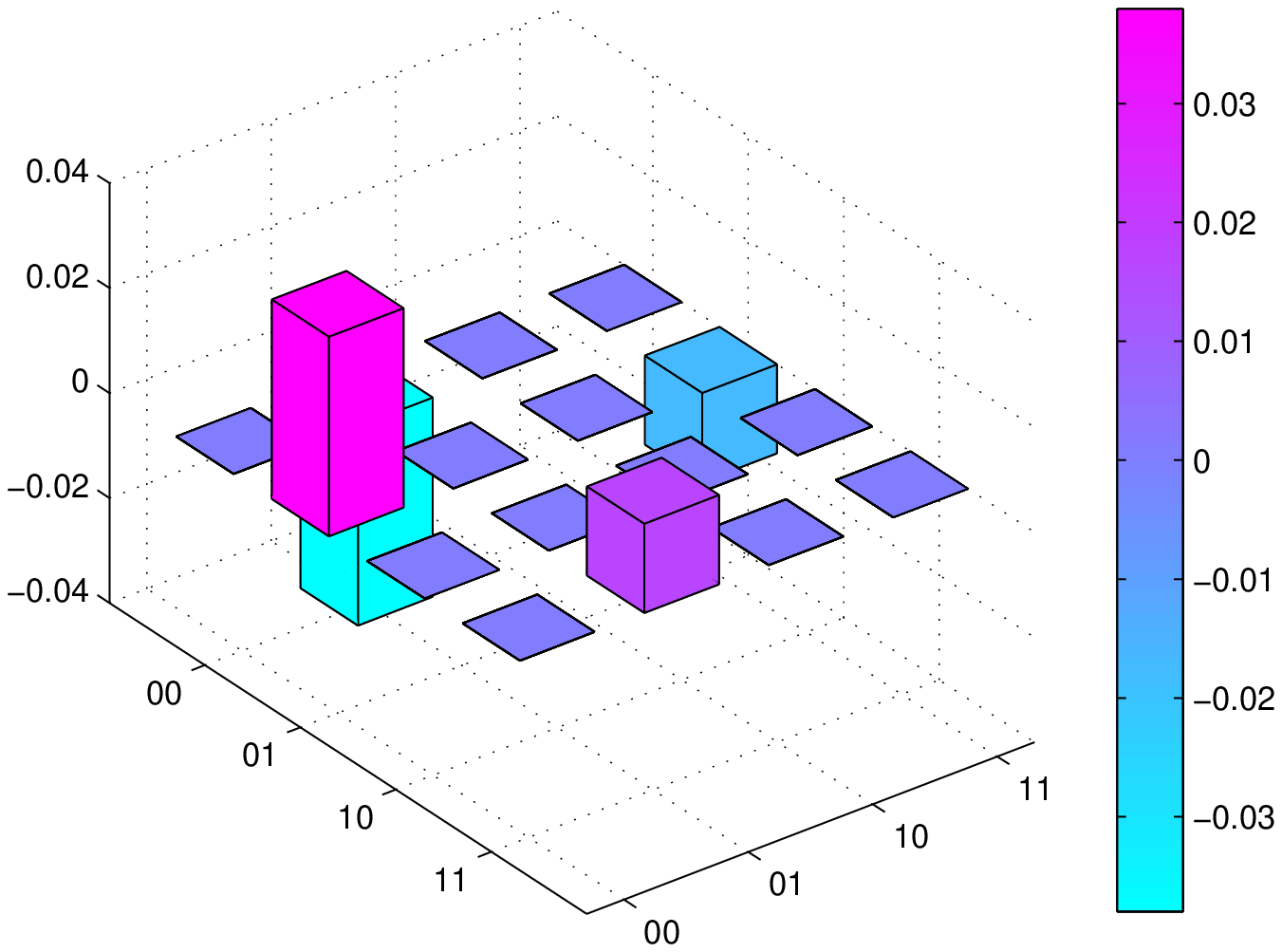}} \\
\subfigure[]{\includegraphics[width=7.0cm]{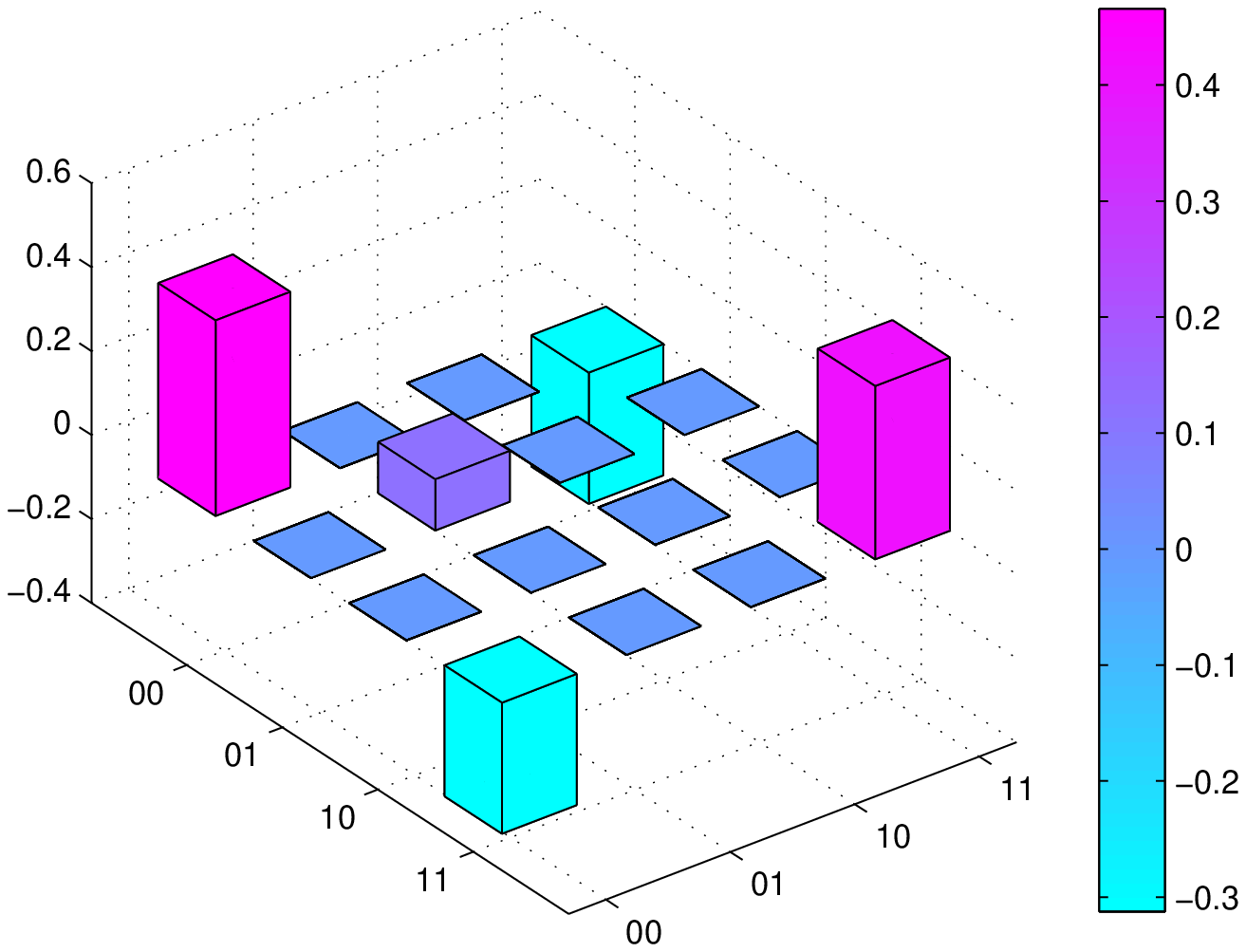}}\hfill
\subfigure[]{\includegraphics[width=7.0cm]{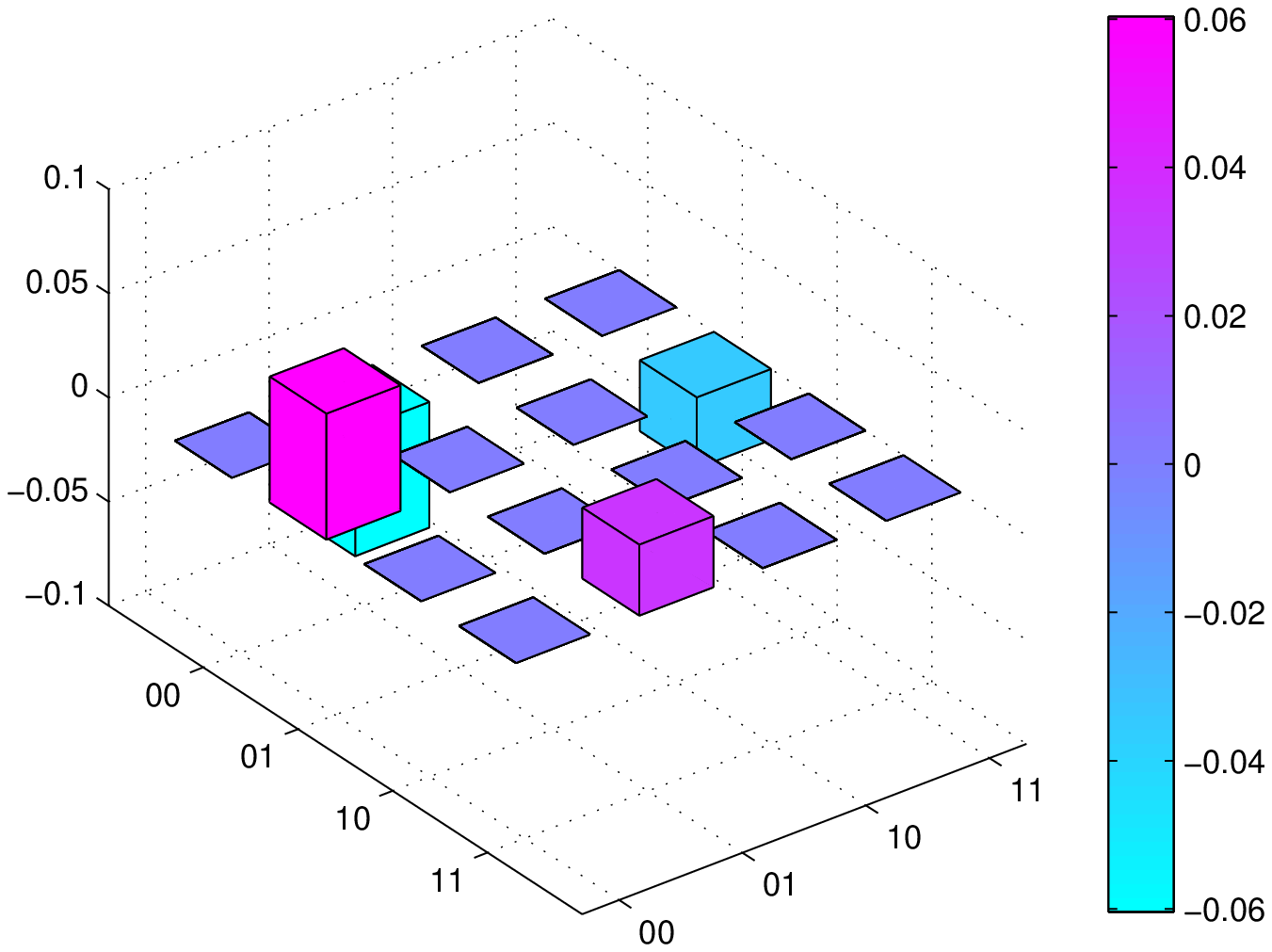}} \\
\subfigure[]{\includegraphics[width=7.0cm]{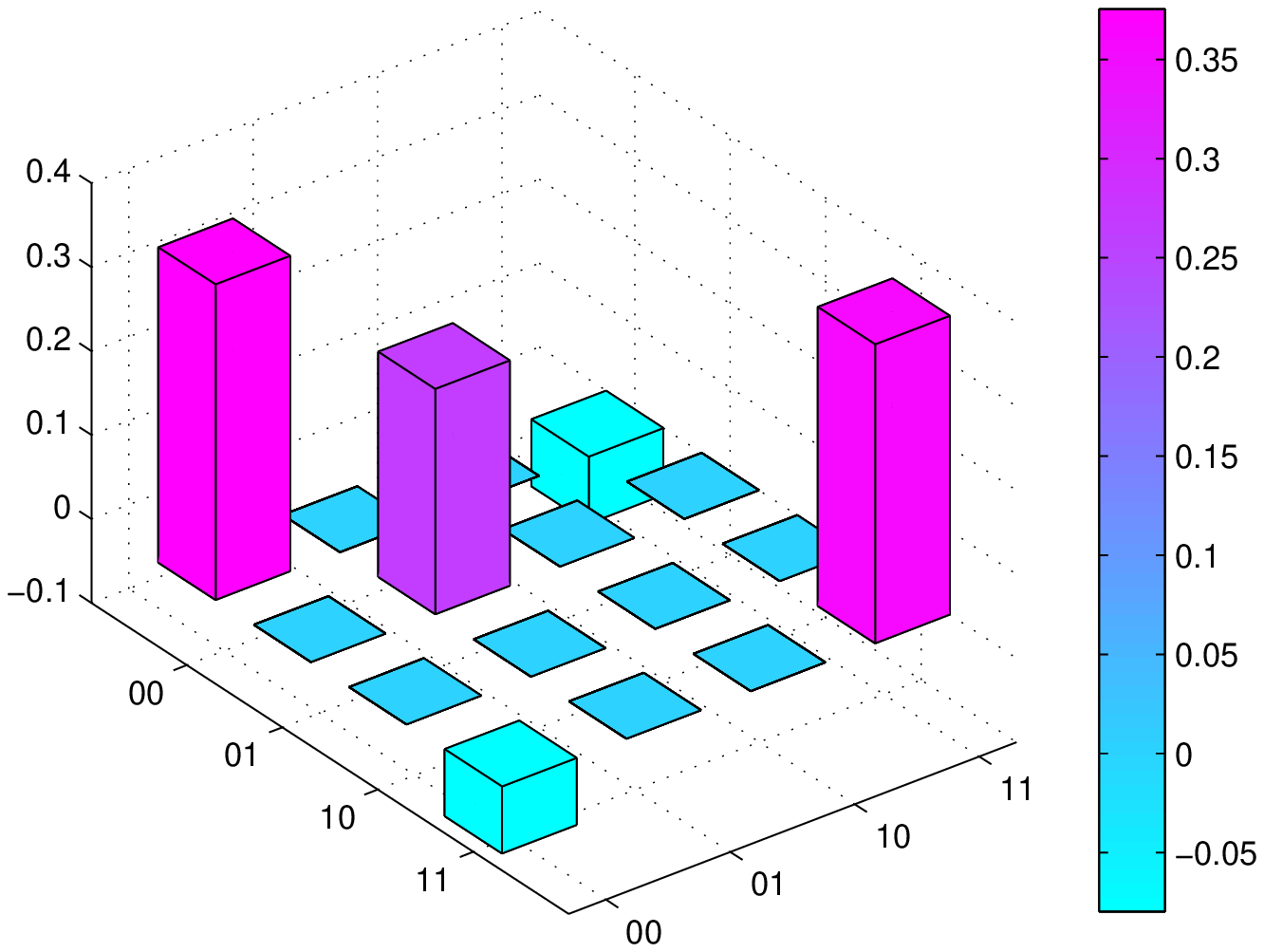}} \hfill
\subfigure[]{\includegraphics[width=7.0cm]{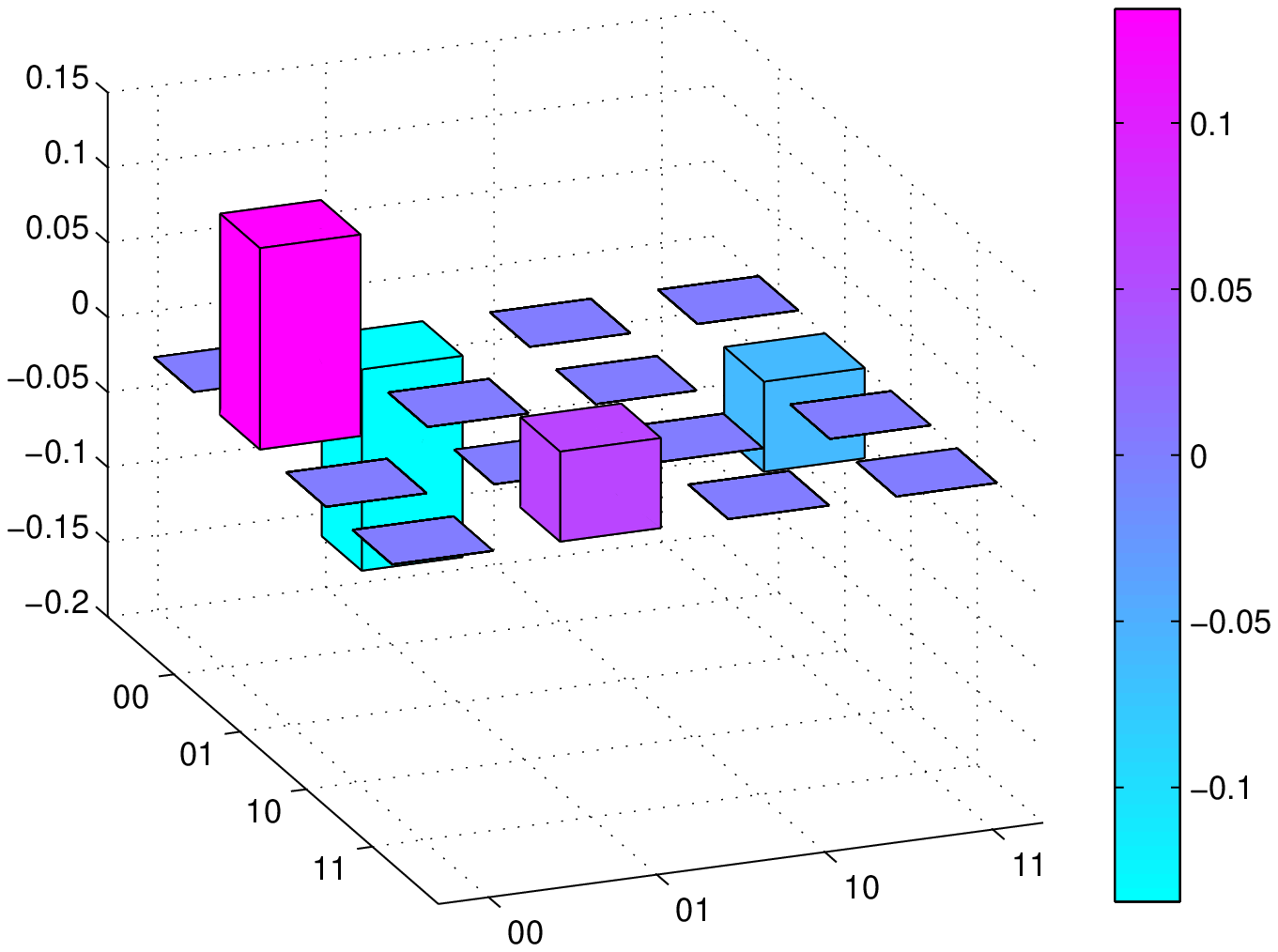}}\\
\caption{(Color online) Real part of the two photon density matrix for (a) the pure state regime,
(c) the Bell state regime  and (e) the mixed state regime and (b), (d), and (f) show the imaginary part of the
density matrix in the respective regimes. The parameters are the same as in Fig.2.}
\label{fig:density matrix}
\end{figure}

\subsection{Numerical results}

We now discuss the numerical results in three regimes of interest: a) the pure state regime 
$\Omega_1< \Omega_2$, b) the Bell state regime, $\Omega_1 \approx \Omega_2$ and c) the mixed state regime $\Omega_1 > \Omega_2$. The reason for this nomenclature becomes clear in our analysis below. All the system parameters may be rendered dimensionless by scaling with  the atomic lifetime $\gamma \approx 1 MHz$ .  In particular we choose the pure state regime corresponding  to  the driving field strengths $\Omega_1 = 3.0$ and $ \Omega_2= 6.0$. Our numerical simulations correspond to the two-photon resonant situation;  therefore, $\Delta_1=\Delta_2=0$. The  upper most level $|3\rangle$ (which may correspond to $5D_{5/2})$ is assumed to be metastable,
 i.e., $\Gamma_3=1.0$ and $\Gamma_2=6.0$.  We further assume the system to be closed.

Fig.2 shows the populations in the $|00\rangle$,
$|01 \rangle$  and the $|11\rangle$ of the two photon state as  functions of time. We have studied the time dependence of the populations in the three regimes mentioned above. Fig.2a corresponds to the pure state regime wherein the population is negligible in  $ |01\rangle$  and is mostly in   $|00\rangle$ with a relatively small admixture of $|11\rangle$.
 Indeed it  is clear from the graph that the purity, $Tr(\rho^2)$,  is quite close to one. The Bell state regime corresponds to 
$\Omega_1=6.0, \Omega_2=6.0 $ and is illustrated in Fig.2b.  Here, the level $|01\rangle$  starts getting populated but remains small compared to the populations in $|00\rangle$ and the $|11\rangle $ states which become comparable to each other. The coherence between $|00\rangle $ and $ |11\rangle$ is also quite large. In  the mixed state regime ($\Omega_1= 6.0, \Omega_2= 3.0$)  shown in  Fig.2c, we  note that all the populations are comparable and the purity is much less than one. Note  that the time evolution in all the regimes attains steady state
 in approximately an atomic lifetime after performing one Rabi cycle. This justifies the single photon approximation approach \cite{single} used here. 

The steady state two-photon density matrix  corresponding to the three regimes is shown in Figs.3a-3f . 
The real part of the density matrix in the pure, the Bell and the mixed state regimes are respectively shown in Figs.3a, 3c and 3e.  Likewise the imaginary part of the density matrix in the three regimes are shown in Fig.3b, 3d and 3f. Note that the third column and the third row are zero everywhere, as implied  by Schmidt decomposition and the dynamics. 
  In the pure state regime (Figs.3a, 3b), even  though most of the population is in the ground state there is a nonzero population in the $|11\rangle$ state and a  nonzero correlation in the $\rho^{\gamma}_{00; 11}$ element.  This is an almost pure state as is clear from Fig.2a ($Tr \rho^2 \approx 1$) and the pure state component is of the form $c_1|00\rangle + ~ c_2|11\rangle;~ c_1 \gg c_2$. The small mixedness is due to the nonzero population in the $|01\rangle$ state. In the Bell state regime  (Figs.3c, 3d), the correlation is strong  and  further the populations in the  $|00\rangle$ and $|11\rangle$ are equal . The  loss of purity is again due to the $|01\rangle$ state. In the mixed state regime (Figs.3e,3f), the steady populations in the three levels are comparable, and the correlations $\rho_{00; 01}$ and $\rho_{01;11}$ compete with the correlation $\rho_{00;11}$. This regime provides the most general two photon mixed state .
  
Having determined the state of the two-photon density matrix we now move on to study the entanglement properties of the two-mode emitted radiation.

\section{Control and Characterization of Entanglement}

Pure state entanglement  may be characterized by various measures such as the Von Neumann entropy of the reduced density matrix, concurrence and negativity. All these measures are equivalent in the sense that they are relative monotones with respect to each other. In contrast,
it is well known that the entanglement of a  mixed state (MSE) cannot be characterized by a single parameter \cite{bennet}.  Generalizations of the above mentioned  pure state measures in terms of a single parameter cease to be relative monotones of each other in the mixed state regime;  this has been  demonstrated explicitly in the case of concurrence and negativity  \cite{grudka}. In  a  detailed and exhaustive analysis,  Peters et. al. \cite{kwiat_03, kwiat_04} 
have analyzed the sensitivity of the state to a number of measures of entanglement for depolarized nonmaximally entangled states such as the Werner states and also the so called maximally entangled mixed states. 
They perform a relative  comparison of six benchmarks of entanglement {\it viz.}, fidelity, trace distance, tangle, linear entropy, concurrence and von Neumann entropy,
and find an imbalance in the values of these quantities, even in situations where it is not expected, clearly signalling that none of these can by itself fully account for MSE.
 Finally, we mention quantum discord \cite{zurek} which again displays the tension  between such definitions of entanglement.

We would like to point out here  that the description of mixed state entanglement by Bhardwaj and Ravishankar \cite{shantanu} addresses these issues
by characterizing MSE in terms of a probability density function (PDF) rather than a single number. This characterization does not invoke any new concept such as entanglement of formation or separability. Instead, it is based on the observation  that a description in terms of mixed states is required when we have an ensemble of quantum systems each of which is in a pure state. A probabilistic description is, therefore, natural. This expectation is vindicated by the fact that while two states which are not equivalent under local operations ($SU(2) \otimes SU(2)$) can have the same value for concurrence, the probability densities are necessarily distinct. In fact, the state can be reconstructed almost uniquely (up to local transformations) by the density. We employ this characterization of MSE in this paper. We briefly describe their formalism:
\subsection{Probability density function for MSE}
The characterization of mixed state is accomplished in two steps. Firstly, the special case when the state is a projection of rank $d$ is considered 
(For a two qubit system,  $d_{max}=4$), $ \rho_d \equiv \frac{1}{d} \mathds{P}_d$, where the operator $\mathds{P}_d$ projects a $d$ dimensional subspace 
$\mathcal{H}_{\mathds{P}_d}$ of the two qubit Hilbert space $\mathcal{H}_4$. We then observe that for all 
$\psi \in \mathcal{H}_d,~ \langle \psi \rangle_{\rho_d} =1$, which means that there is a uniform probability for 
the system to be in any state in the subspace $\mathcal{H}_d$. By the same token, for all states  
$\psi \notin \mathcal{H}_d,~
 \langle \psi \rangle_{\rho_d} =0$, implying that the probability for finding the system in the complementary subspace vanishes identically. 
 Armed with this
 simple basis independent observation, define the probability density function 
\begin{equation}
{\cal P}_{\rho_d}({\cal E})=\frac{\int d{\cal V}_{\mathcal{H}_d}\delta ({\cal E}_{\psi}-{\cal E})}{\int d {\cal V}_{\mathcal{H}_d}}
\end{equation}
where $d {\cal V}_{\mathds{H}_d}$ is the volume measure for the manifold  $\mathcal{H}_{\mathds{P}_d}$. To repeat, the density so obtained is not an artifact of any expansion of $\rho$. We mention that the state may be reconstructed from the resultant PDF, which is  unique --- up to local transformations.

In the case when $\rho$ is not a projection, it is written as a weighted sum of  projection operators $\mathds{P}_d$, satisfying the nestedness condition
$\mathcal{H}_{\mathds{P}_{d-1}} \subset \mathcal{H}_{\mathds{P}_d}$. Consider  the expansion of $\rho$ in its eigenbasis with its eigenvalues written in a non-increasing order, then, 
\begin{equation}
\rho = \sum_{i=1}^4 \lambda_i^{\downarrow} |\psi^i \rangle \langle \psi^i| \equiv \sum_{d=1}^4 {\rm w}_d \mathds{P}_d;~ {\rm w}_d = \frac{\lambda_d -\lambda_{d-1}}{\lambda_1},
\end{equation}
with the restriction that ${\rm w}_4 =\lambda_4/\lambda_1$.
Employing this resolution, the PDF for any state can be written as
\begin{equation}
{\cal P}_{\rho}({\cal E})= \sum_{d=1}^4 {\rm w}_d {\cal P}_{\rho_d}({\cal E}).
\end{equation}
The PDF so defined is invariant under $SU(2)\otimes SU(2)$ transformations. 

It has been shown by Bhardwaj and Ravishankar  that it requires seven independent parameters which characterize the PDF in various subspaces as  follows:
\begin{eqnarray}
    {\cal P}_{\rho_1} \leftarrow {\rm w}_1,{\cal E}_1\nonumber \\ 
    {\cal P}_{\rho_2} \leftarrow {\rm w}_2,  {\cal E}_{cusp}, {\cal E}_{max} \nonumber \\  
 {\cal P}_{\rho_3} \leftarrow {\rm w}_3, {\cal E}_{\perp} \nonumber \\
       {\cal P}_{\rho_3} \leftarrow  {\rm w}_4.  
\end{eqnarray} 
 Recall that  $\rho_d$ is the projection onto the 
$d$ dimensional subspace. In what follows, for pure state entanglement, we shall employ the pure state concurrence, $\mathcal{E} = 2 \vert \langle \psi^{\intercal}\vert \psi \rangle \vert$, where $\vert \psi^{\intercal}\rangle $ is the time reversed state  of $\vert \psi \rangle$.

\subsubsection{PDF for projective states}
Before we proceed to use the above characterization of mixed state entanglement, a brief description of the parameters is in order. 

\noindent (i) {\it  Pure states}: Consider first a state which is a one dimensional projection. Being pure, it has a sharp value of entanglement. The PDF is, accordingly, given by a Dirac delta: 
\begin{equation}
{\cal P}_{\rho_1} = \delta (\mathcal{E} - \mathcal{E}_{\psi});~ \mathds{P}_1 \equiv \vert \psi \rangle \langle \psi \vert. 
\end{equation}

\noindent (ii) {\it Two dimensional projections} $\rho_2$:  The evaluation of PDF for a two dimensional space involves two parameters and is somewhat involved, given its rich structure. We briefly outline the idea.
Consider a state $|\Psi\rangle \in {\cal H}(\mathds{P}_2$) which may, in general,  be written in terms of two orthonormal states $|\chi_1\rangle , |\chi_2\rangle$ as
\begin{equation}\
|\Psi\rangle = \cos\frac{\theta}{2} e^{i \phi/2}|\chi_1\rangle + \sin\frac{\theta}{2}e^{-i\phi/2}|\chi_2\rangle
\end{equation}
where $0 \le \theta \le \pi $ and $ 0 \le \phi ~\textless ~2 \pi$. It follows from elementary algebra that one can always construct a linear superposition which is separable, which can be conveniently be labelled $\vert 00 \rangle$, after  a further local $SU(2) \otimes SU(2)$ transformation. Using this as  a basis vector, its orthogonal state $ x|01 \rangle + y|10\rangle +z|11\rangle$ may be so chosen, by employing the residual $SO(2) \otimes SO(2)$ symmetry that preserves $|00\rangle$,  such that $x,y,z$ are real nonnegative. By the normalization condition, only two of them, say $x,y$ are independent. In short any state which is a two dimensional projection is characterized by two invariant parameters under local transformation.

The mixed state density matrix of such a state (two qubits $A$ and $B$) may also  be written in the form\begin{equation}
\rho^{AB}=(1+\hat \sigma_A\cdot {\bf P}_A +\hat \sigma_B \cdot {\bf P}_B+ (\hat \sigma_A \otimes \hat \sigma_B)\cdot {\bf \Pi})
\end{equation}
Here it may be shown from the equivalence of the atomic and field operators that ${\bf P}_A$ and ${\bf P}_B$ are the polarizations  due to the two transitions of the atomic system; in particular we identify $A \rightarrow1, B \rightarrow 2$.   The invariant parameters $x,y$   may be written in terms of the polarizations as
\begin{eqnarray}
x^2 &= &{ P}_A^2/(1+ \sqrt{ 1- P_A^2- P_B^2)} \nonumber \\
y^2 &= &{ P}_B^2/(1+ \sqrt{ 1- {P}_A^2- { P}_B^2)} \nonumber \\
z^2&=& (1-x^2-y^2)
\end{eqnarray}
where $P_{A(B)}^2= 1-4 det(\rho^{A(B)})$ and $\rho^A= Tr_B ( \rho^{AB})$. The PDF is shown to attain  the  form 
\begin{equation}
{\cal P}_{\rho_2}({\cal E})= \frac{\cal E}{\pi} \int_0^{\pi} \frac{{\rm sin} \theta  {\rm d} \theta}{\sqrt{[{\cal E}^2-{\cal L}^2(\theta)] [{\cal U}^2(\theta)-{\cal E}^2 ]}}.
\end{equation}
Here ${\cal E}$ satisfies the bounds  $0\le {\cal E} < {\cal U} (\theta),~ {\cal L}(\theta) < {\cal E} \le 0$ where
$ {\cal U}(\theta) = \vert z { \sin}\theta + (1- {\cos}\theta (1-z^2)/2 \vert $ and
${\cal L}(\theta) = \vert z {\sin}\theta - (1- {\cos}\theta (1-z^2)/2 \vert $. Note that in the above equation  the Haar  measure for the volume is simply 
that of the two sphere $\mathcal{S}^2$: $ d \mathcal{V}_{\mathcal{H}_d}= {\rm sin}\theta {\rm d}\theta {\rm d}\phi$.

\noindent {\it Three dimensional projections $\rho_3$}: The PDF for three dimensional projections have a more complicated volume measure, but a simpler structure, made possible by the fact that it is uniquely characterized by its dual, the orthogonal (pure) state $|\psi_{\perp}\rangle$. Let its entanglement be $\mathcal{E}_{\perp}$. The PDF for $\rho_3$ is then given by  
\begin{equation}
{\cal P}_{\rho_3}({\cal E})  = \frac{2 \mathcal{E}}{\sqrt{1-\mathcal{E}_{\perp}^2}}\cosh^{-1} \left(\frac{1}{\mathcal{E}_>}\right)
\end{equation}
where $\mathcal{E}_> = \rm{max}(\mathcal{E},~\mathcal{E}_{\perp})$.
As an interesting aside, we note that concurrence and negativity vanish identically for any three dimensional projection.

The four dimensional projection is simply the fully unpolarized system, which  as a universal background. We do not reproduce the curve here since it has zero weight for our system since 
 one of the eigenvalues of the photon density matrix, corresponding to the state $|10\rangle$,  is zero. There is a corresponding reduction in the number of invariant parameters that characterize entanglement. 
 
  An equivalent description to PDF is given by the cumulative distribution function (CDF) defined by  
\begin{equation}
{\cal F}_{\rho}({\cal E})= \int_0^{\cal E} d{\cal E^{\prime}} {\cal P}_{\rho} ({\cal E^{\prime}}),
\end{equation}
We would like to mention here that we have  evaluated  both the probability density function (PDF) and the cumulative distribution function (CDF) for the entanglement as we find that the latter  displays  the dynamical features more clearly. 
\begin{figure}
\includegraphics[width=8.0cm]{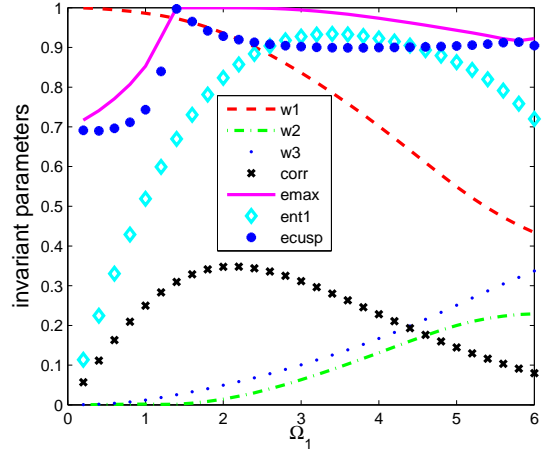}
\caption{(color online) Invariant parameters and correlation as  functions of the ground state excitation energy $\Omega_1$; $\Omega_2= 3.0$. The parameters w1, w2 and w3 correspond to the three weights ${\rm w}_i$,  $'$ent1$'$ = ${\cal E}_1$, $'$ecusp$'$=$ {\cal E}_{cusp}$, $'$emax$'$= ${\cal E}_{max}$ are defined in eq.(24) and $'$corr$'$= $ \rho^{\gamma}_{00;11}$ }
\label{fig:4}
\end{figure}
\begin{figure}
\subfigure[]{\includegraphics[width=8.0cm]{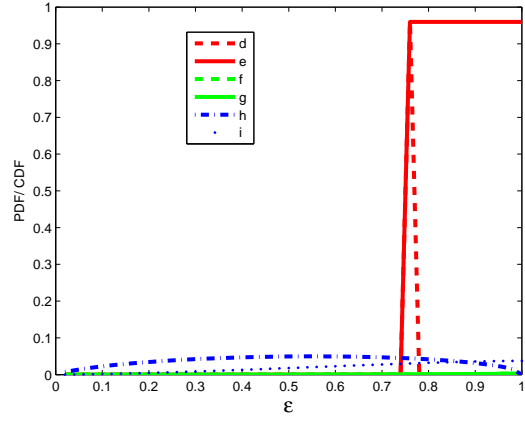}}\\
\subfigure[]{\includegraphics[width=8.0cm]{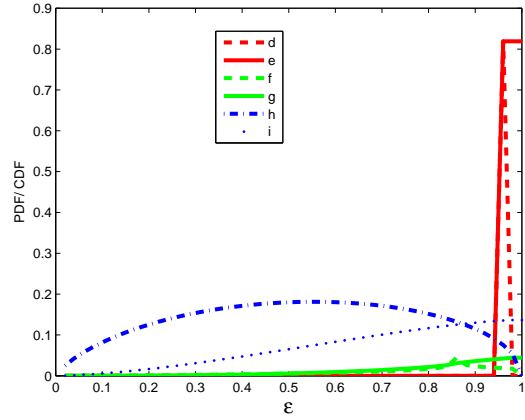}}\\
\subfigure[]{\includegraphics[width=8.0cm]{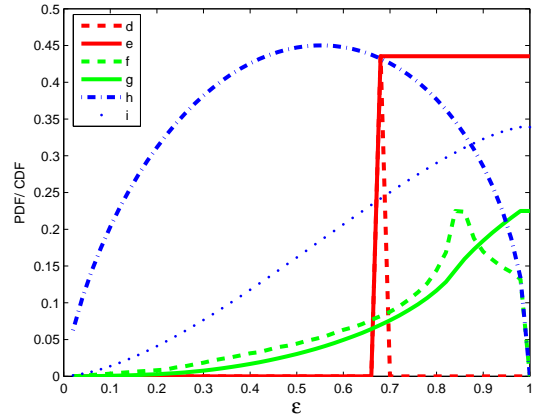}}
\caption{(Color online) Probability distribution and the cumulative distribution of entanglement  corresponding to  one, two and three dimensional subspaces. for the a) pure state, b) the Bell state and c) the mixed state regimes. Legend:( d) probability distribution function for one dimensional sub-space and (e) the corresponding cumulative distribution, (f) PDF for two-dimensional sub-space and (g) its CDF, (h) PDF corresponding to the three dimensional subspace and (i) its corresponding CDF.}
\label{fig:5}
\end{figure}
\subsection{Numerical Results and Discussion}

We have evaluated the steady state values of the six invariant parameters characterizing the mixed state entanglement distribution, {\it viz.}, the three weights  ${\rm w}_1, {\rm w}_2, {\rm w}_3$, the  entanglement of the pure state component ${\cal E}_1$, and the two entanglement parameters 
${\cal E}_{cusp}, {\cal E}_{max}$ corresponding to the two dimensional projection. The parameters $\rm{ w}_4,~{\cal E}_{\perp}$  vanish identically. The parameters ${\cal E}_{max} $,
 ${\cal E}_{cusp}$ are determined from $\rho_2$ (see Eq. 28) by
\begin{eqnarray}
{\cal E}_{max} &=& xy+ \sqrt{z^2+ x^2 y^2}\nonumber \\
{\cal E}_{cusp} &=& z^2/{\cal E}_{max}.
\end{eqnarray}

The three interesting regimes, $\Omega_1~ \textless ~ \Omega_2$, $\Omega_1~ \approx ~ \Omega_2$, and $\Omega_1~ \textgreater ~ \Omega_2$, create the two photon mode system in different states. In the first regime the state has a high degree of purity, but little entanglement. In the second regime, most appropriate for fine tuning, the system loses purity to an extent, but gains in entanglement, as indicated in Figs.2 and 3. The last regime destroys both purity and entanglement.  Fig.4 shows the variation of these parameters as  functions of the Rabi frequency of the ground state excitation $\Omega_1$, with $\Omega_2 =3\gamma$, where all the three regimes show up. But beyond this mildly interesting qualitative feature, it is  clear that in much of the range,  the individual parameters by themselves give  little information on the nature of entanglement  reflecting the fact that no single parameter can completely capture all the aspects of mixed state entanglement.  In contrast, as we show below, the probabiity distribution characterized by these invariant parameters sheds light on not only the nature of mixedness of the state, but also the nature of entanglement without any ambiguity. This description seems to overcome the difficulties raised in references \cite{kwiat_03,kwiat_04}.
In Fig.5 we show the PDF  and the CDF  of entanglement corresponding to each of  one, two and three dimensional subspaces for the three regimes of interest.
It is clear from Fig.5a  which corresponds to the pure state regime, the probability distribution has a dominant contribution of $\approx 95 \%$  due to the one dimensional subspace. The corresponding state is not fully entangled, with  ${\cal E} \approx 0.7$ . Recall that for a  Bell  state  ${\cal E}=1$.  As to the precise nature of the state, one can infer from Fig.3a that the state must be  a
superposition of the $|00\rangle$ and $|11\rangle$ states with an excess of population in the $|00\rangle$ state. This is further substantiated by the strong correlation as shown by the $\rho_{00;11}$ element.  The deviation from purity is due to the unequal populations in the $|00\rangle $ and the $|11\rangle$ state. The two and three dimensional contributions are negligibly small in this case.
In the Bell state  regime, Fig.5b, there is a dominant contribution from the pure state even though the probability decreases to $\approx 80\%$. Note however, that the state in this regime is closer to a Bell state since ${\cal E}$ is close to one. There is a small contribution from the three dimensional subspace.
In the mixed state regime Fig.5c there is a comparable contribution from all the subspaces. The pure state probability has however reduced to less than $50\%$ and also ${\cal E} \approx 0.65$. The nature of mixedness in this regime may be understood by looking at Fig.3c and Fig.5c. Even though the populations in $|00\rangle$ and $|11\rangle$ are nearly equal, the correlation $\rho_{00;11}$ is much weaker. On the other hand the correlation $\rho_{00;01}$  is stronger implying that the superposition of the $|00\rangle$ and $|01\rangle$ is competing. Since both the correlations are comparable the states belong to either two or three dimensional subspace. 

The equivalence of the photon and atom density matrix suggests that the correlation $\rho^{\gamma}_{00;11}(t)$ is proportional to the atomic coherence $\rho^A_{13}(t-r/c)$. Note that maximum entanglement of the two photon occurs when this correlation and hence the atomic coherence is strong. This clearly demonstrates the fact that the quantum correlations of the atomic system manifest as strong correlations in the emitted radiation as well. In the mixed state regime the single and two photon coherences compete and results in a weaker entanglement of the two photons. Thus the control of entanglement/purity is achieved by controlling the atomic coherences.

To conclude, we have shown that it is possible to determine the two photon density matrix, in its Fock space, by a combination of observations that 
include measurements on the atomic system that generate the photons and the radiation itself. By a proper choice of the two driving field strengths, the 
radiation emitted can either be prepared as a pure state, but with little entanglement, or with a reasonable overlap with the Bell state, but with some mixedness. Both the regimes are of interest. The former is eminently suited for the preparation of fiducial states as in applications such as the
 Deutsch algorithm. The latter regime, where highly entangled states are produced is useful when initial entanglement is required, as in the case of teleportation. The third regime, where $\Omega_1 \gg \Omega_2$ seems unsuited for applications; on the other hand, it provides  mixed states exhibiting a rich variety of entanglement which cannot be captured in terms of a simple parameter. It should be of interest to extend this study to multi photon production and its characterization in terms of purity and entanglement.

\acknowledgements
One of us (SNS) thanks Usha Devi for useful discussions. SNS would like to thank the Department of Science and Technology, India for financial support under the WOS-A scheme.


\begin{thebibliography}{100}

\bibitem{reviews} A. I. Lvovsky and M. G. Raymer, Rev. Mod. Phys. {\bf 81} 299 (2000);
G. M. D'Ariano, M. G. A. Paris and M. F. Sacchi, Advances in Imaging and Electron Physics {\bf 128} p205 (2003); Raymer, M. G., and M. Beck, 2004, in Quantum State Estimation, edited by M. Paris and J. Rehacek , Lecture Notes in Physics Vol. 649 (Springer, Berlin), p. 235.

\bibitem{Kwait_pra} D. F. V. James , P. G. Kwait, W. J. Munro and A. G. White, Phys. Rev. {\bf A 64} 052312 (2001).

\bibitem{kumar} Michael Vasilyev, Sang-Kyung Choi, Prem Kumar, and G. Mauro D'Ariano
Phys. Rev. Lett. 84, 2354 (2000).

\bibitem{vogel} K. Vogel and H.Risken, Phys.Rev. {\bf A 40} 2847 (1989).

\bibitem{raymer} D. T. Smithey, M. Beck, M. G. Raymer, and A. Faridani, Phys. Rev. Lett. 70, 1244 (1993). 

\bibitem{wodkiewicz} Konrad Banaszek and Kryzysztof Wodkiewicz, Phys. Rev. Lett. 76, 4344 (1996).

\bibitem{lvovsky1} A. I. Lvovsky, H. Hansen, T. Aichele, O. Benson, J. Mlynek, and S. Schiller, Phys. Rev. Lett. 87, 050402 (2001).

\bibitem{lvovsky2} S. A. Babichev, J. Appel, and A. I. Lvovsky, Phys. Rev. Lett. 92, 193601 (2004).

\bibitem{D'ariano} G M D'Ariano et al, J. Opt. B: Quantum Semiclass. Opt. 5 77-84   (2003).

\bibitem{sns} S. N. Sandhya, Phys. Rev. A 76, 013802 (2007) .

\bibitem{zubairy}Fu-li Li, Han Xiong, and M. Suhail Zubairy, Phys. Rev. A 72, 010303 (2005).

\bibitem{scully} Hua-Tang Tan, Shi-Yao Zhu, and M. Suhail Zubairy, Phys. Rev. A 72, 022305 (2005).

\bibitem{shantanu} Shanthanu Bhardwaj and V. Ravishankar, Phys. Rev. A 77, 022322 (2008).

\bibitem{single} B. R.Mollow, Phys. Rev. {\bf A 12} 1919 (1975).

\bibitem{clauser} J. F. Clauser, Phys. Rev. D 9, 853 (1974) .

\bibitem{kimble}H. J. Kimble and L. Mandel, Phys. Rev. A 13, 2123 (1976).

\bibitem{loudon} R. Loudon, Rep. Prog. Phys. 43, 58 (1980).

\bibitem{banacloche} Julio Gea-Banacloche, Yong-qing Li, Shao-zheng Jin, and Min Xiao, Phys. Rev. A 51, 576 (1995).

\bibitem{glauber}U. M. Titulaer and R. J. Glauber, Phys. Rev. 145, 1041 (1966).

\bibitem{knight} A. Ekert and P. L. Knight, Am. J. Phys 63 (5), 415. (1995).

\bibitem{cohen} C. Cohen-Tannoudji, J. Dupont-Roc, and G. Grynberg, Atom-Photon Interactions (Wiley, New York, 1992) p380.

\bibitem{scully1} M. O. Scully and M. S. Zubairy, Quantum Optics (Cambridge University Press, Cambridge, 1997).

\bibitem{mandel} L. Mandel, E. Wolf Optical Coherence and Quantum Optics (Cambridge 1995).

\bibitem{lax} M. Lax, Phys. Rev. 129, 2342 (1963); ibid., 157, 213 (1967).

\bibitem{gsa} G. S. Agarwal, Phys. Rev. A 34, 4055 (1986).

\bibitem{bennet} C. H. Bennett, D. P. DiVincenzo, J. A. Smolin, and W. K.Wootters, Phys. Rev. A 54, 3824 (1996).

\bibitem{grudka} A. Miranowicz and A. Grudka,  J. Opt. B: Quantum semiclassical Opt. 6, 542 (2004); see also  
  Dagmar Dru$\beta$, J. Math. Phys. 43, 4237 
 2002 for a critique of measures of entanglement.


\bibitem{kwiat_03} Tzu-Chieh Wei, Kae Nemoto, Paul M. Goldbart, Paul G. Kwiat, William J. Munro, and Frank Verstraete, Phys. Rev. A 67, 022110 (2003) .

\bibitem{kwiat_04} Nicholas A. Peters, Tzu-Chieh Wei, and Paul G. Kwiat, Phys. Rev. A 70, 052309 (2004)

\bibitem{zurek} H. Ollivier and W. H. Zurek, Phys. Rev. Lett., 88, 017901 (2002)



\end{thebibliography}
\end{document}